\documentclass[showpacs,preprintnumbers,showkeys,superscriptaddress]{revtex4}
\usepackage{amssymb}
\usepackage{graphicx}
\usepackage{dcolumn}
\usepackage{bm}
\usepackage{appendix}

\def\eqref#1{Eq.~(\ref{eq:#1})}

\begin{document}

\title{Equivalent Representations of Collective Hamiltonian \\
and Implication on Generalized Density Matrix Method}
\author{L. Y. Jia}  \email{jial@nscl.msu.edu}
\affiliation{National Superconducting Cyclotron Laboratory and
Department of Physics and Astronomy, Michigan State University, East
Lansing, Michigan 48824, USA }
\author{V. G. Zelevinsky}
\affiliation{National Superconducting Cyclotron Laboratory and
Department of Physics and Astronomy, Michigan State University, East
Lansing, Michigan 48824, USA }

\date{\today}

\begin{abstract}

We discuss equivalent representations of the collective/bosonic
Hamiltonian in the form of Taylor expansion over collective
coordinate and momentum. Different expansions are equivalent if they
are related by a transformation of collective variables. The
independent parameters in the collective Hamiltonian are identified,
which are much less in number than it appears. In this sense, the
microscopic generalized density matrix method fixes the collective
Hamiltonian completely \cite{Jia}, which seems to solve the old
problem of microscopic calculation of the collective Hamiltonian.

\end{abstract}


\vspace{0.4in}

\maketitle

Recently the generalized density matrix (GDM) method \cite{Jia} was
used to calculate microscopically the collective Hamiltonian. There
we mention that the collective Hamiltonian is not completely fixed;
we only find one constraint in each even order of anharmonicity.

In this work we show that these constraints actually fix the
collective Hamiltonian completely. The collective Hamiltonian is
first constructed as a Taylor expansion of collective coordinate and
momentum, keeping all terms with the right symmetry. However, lots
of different expansions are actually equivalent, related by
transformations of the collective variables. The number of
independent parameters in the collective Hamiltonian is much less
than it appears. Those constraints in Ref. \cite{Jia} are just
enough to fix these independent parameters.

The Taylor expansion of the collective Hamiltonian is
\begin{eqnarray}
H = \frac{\omega^2}{2} \alpha^2 + \frac{1}{2} \pi^2 +
\frac{\Lambda^{(30)}}{3} \alpha^3 + \frac{\Lambda^{(12)}}{4}
\{\alpha, \pi^2\} + \frac{\Lambda^{(40)}}{4} \alpha^4 +
\frac{\Lambda^{(22)}}{8} \{ \alpha^2 , \pi^2 \} +
\frac{\Lambda^{(04)}}{4} \pi^4 + \Lambda^{(50)} \frac{\alpha^5}{5} +
\ldots  \label{H_b}
\end{eqnarray}
Estimates of quite general type show that $\Lambda^{(mn)} \sim
\Omega^{-(m+n-2)/2}$ (see Ref. \cite{Zele_estimate}), where $\Omega$
is the collectivity factor. The Lipkin model and quadrupole plus
pairing model confirm these estimates. We do a transformation of the
collective variables $(\alpha, \pi) \rightarrow (\bar{\alpha},
\bar{\pi})$,
\begin{eqnarray}
\alpha = \sum_{mn} x^{(mn)} \frac{\{\bar{\alpha}^m,\bar{\pi}^n\}}{2
~ m! ~ n!} ~,~~~ \pi = \sum_{mn} y^{(mn)}
\frac{\{\bar{\alpha}^m,\bar{\pi}^n\}}{2 ~ m! ~ n!} ~,
\label{col_tran}
\end{eqnarray}
while keeping the commutation relation
\begin{eqnarray}
[\alpha , \pi] = [\bar{\alpha} , \bar{\pi}] = i .  \label{comu}
\end{eqnarray}
In Eq. (\ref{col_tran}) the summation index $m \ge 0$, $n \ge 0$.
$x^{(00)} = y^{(00)} = 0$ because a trivial translation of origin is
not interesting [there are no linear terms in the collective
Hamiltonian (\ref{H_b})]. $x^{(mn)}$ and $y^{(mn)}$ vanish for odd
and even $n$, respectively, because of the wrong time-reversal
symmetry. $x^{(10)}$ and $y^{(01)}$ are fixed to be $1$ because we
do not want trivial rescaling by a numerical factor ($\Lambda^{(02)}
= 1$). The coefficients $x^{(mn)} \sim \Omega^{-(m+n-1)/2}$,
$y^{(mn)} \sim \Omega^{-(m+n-1)/2}$; thus, transformations
(\ref{col_tran}) do not change the dependence on $\Omega$ of
$\Lambda^{(mn)}$ in Eq. (\ref{H_b}). Using Eqs. (\ref{col_tran}) and
(\ref{comu}) we have
\begin{eqnarray}
[ \alpha , \pi ] = i \cdot \sum_{rsmn} x^{(mn)} y^{(r-m,s-n)} \cdot
\frac{[m (s-n) - n (r-m)]}{2 ~ m! ~ n! ~ (r-m)! ~ (s-n)!} \cdot \{
\bar{\alpha}^{r-1} , \bar{\pi}^{s-1} \}  ,
\end{eqnarray}
where in the coefficient of $\{ \bar{\alpha}^{r-1} , \bar{\pi}^{s-1}
\}$ we keep only the leading terms in $1/\Omega$, that is, terms
$\sim \Omega^{-(r + s - 2)/2}$. The summation index $r \ge m$, $s
\ge n$. If $r = 0$ or $s = 0$, the numerator in the fraction $[m
(s-n) - n (r-m)]$ vanishes; thus, $r \ge 1$ and $s \ge 1$. $s$ is
odd, otherwise $x^{(mn)} y^{(r-m,s-n)}$ vanishes. The $r = s = 1$
term gives correctly $i$. Terms with $r + s \ge 3$ and an odd $s$
should vanish:
\begin{eqnarray}
0 = \sum_{mn} x^{(mn)} y^{(r-m,s-n)} \cdot \frac{ m (s-n) - n (r-m)
}{2 ~ m! ~ n! ~ (r-m)! ~ (s-n)!}  .    \label{constraint}
\end{eqnarray}
These relations constrain $x^{(mn)}$ and $y^{(mn)}$ in the
transformations (\ref{col_tran}).

Different expansions (\ref{H_b}) of the collective Hamiltonian are
equivalent if they are related by transformations (\ref{col_tran}).
Let us identify the independent parameters in the collective
Hamiltonian (\ref{H_b}). In the harmonic order, the transformations
(\ref{col_tran}) do not change the harmonic terms
$\frac{\omega^2}{2} \alpha^2 + \frac{1}{2} \pi^2$; thus, there is
one independent parameter $\omega^2$. In the cubic order, the
transformations (\ref{col_tran}) with nonzero $x^{(20)}$,
$x^{(02)}$, and $y^{(11)}$ influence $\Lambda^{(30)}$ and
$\Lambda^{(12)}$ (through the harmonic terms $\frac{\omega^2}{2}
\alpha^2 + \frac{1}{2} \pi^2$); and there is one constraint
(\ref{constraint}) with $(rs) = (21)$. Thus, $\Lambda^{(30)}$ and
$\Lambda^{(12)}$ can be set to zero and there is no independent
parameter in this order. In the quartic order, the transformations
(\ref{col_tran}) with nonzero $x^{(30)}$, $x^{(12)}$, $y^{(21)}$,
and $y^{(03)}$ influence $\Lambda^{(40)}$, $\Lambda^{(22)}$, and
$\Lambda^{(04)}$; and there are two constraints (\ref{constraint})
with $(rs) = (31)$ and $(13)$. Thus, there is one independent
parameter; we can for example choose it to be $\Lambda^{(40)}$, and
set $\Lambda^{(22)}$ and $\Lambda^{(04)}$ to zero. This counting
continues to higher orders of anharmonicities. There is one
independent parameter in each even order of anharmonicity (for
example we can choose it to be $\Lambda^{(n0)}$); and there are no
independent parameters in odd orders. In summary, the independent
parameters in the collective Hamiltonian (\ref{H_b}) can be
identified in the following form:
\begin{eqnarray}
H = \frac{1}{2} \pi^2 + V(\alpha^2) ~,~ V(\alpha^2) = \omega^2
\frac{\alpha^2}{2} + \Lambda^{(40)} \frac{\alpha^4}{4} +
\Lambda^{(60)} \frac{\alpha^6}{6} + \Lambda^{(80)}
\frac{\alpha^8}{8} + \ldots    \label{H_free_para}
\end{eqnarray}

In Ref. \cite{Jia} we show that the GDM method fixes all the
independent parameters in Eq. (\ref{H_free_para}); thus, it fixes
the collective Hamiltonian completely. In practical application, Eq.
(\ref{H_free_para}) may not be a convenient choice for the
independent parameters, which means solving the equations of motion
in the GDM method to infinitely high orders. Alternatively, we can
pick up a certain number (labeled $l$) of terms in Eq. (\ref{H_b}),
putting other terms to zero; in other words, we \emph{assume} that
the original fermionic Hamiltonian can be mapped onto a collective
Hamiltonian with these $l$ terms. Then in the GDM method we need to
solve the equations of motion up to the order of $2 l$th
anharmonicity, in order to get $l$ constraints. If the above
\emph{assumption} is good, constraints from orders higher than $2
l$th should be satisfied (approximately) identically.

We are testing the above idea in the Lipkin model. There the exact
solution is known; in Eq. (\ref{H_b}) there are only three nonzero
terms: $\omega^2$, $\Lambda^{(40)}$, and $\Lambda^{(04)}$. Thus in
the GDM method we need to go up to the sixth order to fix them.
Constraints from the eighth order and higher should be satisfied
identically.

This work is supported by the NSF grants PHY-0758099 and
PHY-1068217.


\begin{thebibliography}{50}



\bibitem{Jia} L. Y. Jia, {\tt
arXiv:1107.2184v1 [nucl-th]}, accepted to Phys. Rev. C.

\bibitem{Zele_estimate}   V. G. Zelevinsky, Int. J. Mod. Phys. {\bf E2}, 273
(1993).



\end{thebibliography}
\end{document}